\documentclass[11pt,twoside]{article}
\usepackage{newpasp,epsf}
\def\edcomment#1{\iffalse\marginpar{\raggedright\sl#1\/}\else\relax\fi}
\def\gsim{\raise0.3ex\hbox{$>$}\kern-0.75em{\lower0.65ex\hbox{$\sim$}}}
\def\lsim{\raise0.3ex\hbox{$<$}\kern-0.75em{\lower0.65ex\hbox{$\sim$}}}
\reversemarginpar

\begin{document}
\title{Structure Shocks as a Source of Cosmic Rays in Clusters}
\author{T. W. Jones}
\affil{University of Minnesota, Department of Astronomy, 116 Church St. SE,
Minneapolis, MN 55455, USA}
\author{Francesco Miniati}
\affil{Max-Planck-Institut f\"ur Astrophysik, Karl-Schwarzschild-Str. 1, D-85741
Garching, Germany}
\author{Dongsu Ryu}
\affil{Department of Astronomy and Space Science, Chungnam National
University, Daejun 305-764, Korea}
\author{Hyesung Kang}
\affil{Department of Earth Science, Pusan National University,
Pusan 609-735, Korea}
\author{Eric J. Hallman}
\affil{University of Minnesota, Department of Astronomy, 116 Church St. SE,
Minneapolis, MN 55455, USA}

\begin{abstract}
Shocks are a ubiquitous consequence of cosmic structure formation,
and they play an essential role in heating galaxy cluster
media. Virtually all of the gas in
clusters has been processed by one or more shocks
of at least moderate strength. 
These are collisionless shocks, so likely sites
for diffusive shock acceleration of
high energy particles. We have carried out numerical
simulations of cosmic structure formation
that directly include acceleration and
transport of nonthermal protons, as well as primary and secondary
electrons. Nonthermal emissions have also been
computed from the resulting particle spatial
and energy distributions. Here we outline some
of our current findings, showing that
nonthermal protons may contribute a significant
pressure in cluster media, and that expected radio,
X-ray and $\gamma$-ray emissions from these
populations should be important cluster diagnostics.
\end{abstract}

\section{Introduction}

At least two lines of reasoning lead us to examine
the properties of energetic particle acceleration at structure formation shocks. 
First, virtually
all the gas in filaments, groups and clusters\footnote[1]{To
simplify the discussion below we do not distinguish groups from
clusters.} has at some time passed through one, or 
more probably, several structure shocks (e.g., Quilis et al 1998; Cen \& Ostriker 1999;
Miniati et al. 2000). Structure shocks involve
very diffuse plasmas, so will be ``collisionless''.  Thus, they are
likely to accelerate high energy particles (call them ``cosmic rays'' or CRs)
through the so-called ``diffusive
shock acceleration'', provided there is a weak magnetic 
field present (e.g., Blandford \& Eichler 1987).
As discussed by Blasi at this meeting and outlined briefly below,
clusters should be good reservoirs of CR protons, so that over
time they will accumulate, possibly even to a level where 
they can contribute significantly to the cluster pressure (e.g., Berezinsky et al 1997).
CRs have a softer equation of state than nonrelativistic thermal plasma,
but CR protons are effectively immune to nonadiabatic cooling. 
Consequently, they can
change the thermodynamic properties of the ICM. In that
event it becomes important to include them in consideration of
cluster dynamics, especially in cooling flows
(see, e.g., the contribution by Ryu et al in these proceedings).

The second rationale for understanding particle acceleration at
structure shocks comes from diffuse nonthermal emissions
seen in at least some clusters. 
The most compelling such evidence, known for some time, is the
existence of diffuse 
cluster radio halos and 
so-called ``radio relic'' sources (e.g., Feretti \& Giovannini 1996;
Feretti 1999). The detailed
properties of these two classes of radio source are 
different in some respects, such as polarization and location inside the clusters, 
but both involve
substantial volumes in their host clusters. The radio halos tend to be centered 
on the cluster cores and are unpolarized, while the relics are most likely to be found on
the perimeters of clusters and can be highly polarized. 
Both types of radio sources result from
synchrotron radiation by substantial populations of $\gsim$GeV electrons. 

As discussed at this meeting by Fusco-Femiano, X-ray
observations now show convincing evidence for
diffuse, nonthermal hard X-ray emission in at least Coma and A2256 (e.g.,
Rephaeli et al 1999; Fusco-Femiano et al 2000). 
Again, this implies nonthermal electron populations. Here, however,
as discussed by several other speakers at this workshop, the origin of that
emission and the energy of the electrons are less certain. If the
emission is nonthermal bremsstrahlung the electrons are only a little 
more energetic than the thermal electrons responsible for the soft X-ray
continuum. On the other hand, one of the
prime candidates for the hard X-ray excesses is inverse-Compton scattered
CMB photons, again involving roughly GeV electrons. 
So far, the evidence requires only nonthermal electrons, since $\pi^0$ decays
coming from inelastic p-p collisions have not yet been detected.  However,
for particle acceleration in normal plasmas we should expect
energetic hadrons as well. If the accelerators behave similarly to
galactic CR accelerators the energy carried by hadronic CRs
is likely to be one to two orders of magnitude greater than for
electrons, in fact. 

To facilitate what follows it may be useful here to review
very quickly the key issue of CR longevity in clusters, since that
largely controls the needs for extended accelerators of CRs. First, it
has long been noted that electrons responsible for the observed
radio halos have such short lifetimes
to radiative losses that they cannot possibly fill a cluster from
a single point source (e.g., Jaffe 1977). Supposing for example, that
the cluster magnetic field is $1 \mu$G, then electrons radiating
at 1 GHz have Lorentz factors, $\gamma \sim 2 \times 10^4$. For Lorentz factors
above a few hundred the dominant energy losses will come from
inverse-Compton scattering in this case (e.g., Sarazin 1999), 
leading to lifetimes $t_r \sim 4\times 10^{12} \gamma^{-1}$ yrs,
or about $2\times 10^8$ yrs in this case.
It is simple to show that if we fix the observed radio frequency at 1 GHz,
this is about the maximum
lifetime of the relevant radiating electrons against combined inverse-Compton
and synchrotron radiation.

Although in free flight relativistic electrons could cross a cluster
in a few million years, diffuse radio emissions from the clusters
and Faraday rotation through some clusters 
reveal the clear presence of weak magnetic
fields (see the contributions by Feretti,  Clarke, and Kronberg at this
meeting, for example). The observations point to a tangled, 
perhaps turbulent field. MHD fluctuations will severely
restrict their propagation. As an example consider Bohm diffusion,
corresponding to scattering on saturated field fluctuations,
so that a particle mean free path
approximates the particle gyro radius; that is, $D_B = \frac{1}{3} c r_g$,
for relativistic particles. This would allow GeV particles
to diffuse less than a kiloparsec in a Hubble time. Bohm diffusion is a
limiting case, and advective
motions will surely carry CR electrons farther than that by a large factor even in
the much smaller radiative lifetime of the electrons.
But, the very small range of these electrons
remains valid under any reasonable set of circumstances. The main
consequence of this result is that CR electrons responsible for
observed diffuse cluster radio and probably X-ray and 
predicted $\gamma$-ray emissions must be continuously replenished somehow.
The observed hard X-ray luminosity of Coma exceeds $10^{43}$erg s$^{-1}$,
placing a lower bound on the replenishment rate in that cluster.

Note as well that hadronic CRs, and protons, in particular, are confined similarly by
MHD wave scattering. In that case, on the other hand, radiative losses are
negligible up to extremely high energies, as are losses due to inelastic
collisions with the CMB and the cluster thermal plasma (e.g., Berezinsky et al 1997).
So, the key consequence is that CR protons below about $10^{15}$eV = 1 PeV
will be essentially locked to the ICM forever, once they are introduced.

The importance of shock heating to the ICM makes diffusive shock acceleration
(DSA) an immediate candidate
for production of cluster CRs, since DSA can transfer some tens of percents
of the energy dissipated at the shock to the
CR population (e.g., Blandford \& Eichler 1987). For a typical cluster shock that would
correspond to a CR  energy input rate $\sim 0.1 \rho u^3 R^2 \sim 10^{45}$ erg/sec
(e.g., Jones et al 2001).
Over a Hubble time this could amount to as much as $10^{63}$ergs of CR
energy.
We emphasize from the start that our aim is not to argue for structure
shocks as the only source of diffuse CRs in clusters, but that it
is one very
likely to be there and to be important. Others have estimated
that radio galaxies can contribute a similar CR energy flux (e.g., En$\ss$lin et al 1997).
Intense starburst activity during galaxy formation has also been proposed (e.g., V\"olk et al 1996). Turbulent
acceleration in the ICM may also play a significant role, especially
as a means to reaccelerate CRs introduced by some other means (e.g., Jaffe
1977; Eilek \& Wetherall 1999; Brunetti et al 2001).

In the remainder 
of this paper we will outline the role of structure shocks in determining the 
conditions in the ICM, review briefly the relevant properties of
shock acceleration physics and discuss the
results of some numerical simulations of structure formation
that include treatment of CR acceleration via DSA, followed by
advective transport and relevant energy loss mechanisms. Finally,
since the ultimate test of such calculations is their match to
observed cluster properties, we have computed ``synthetic observations''
of nonthermal emissions from simulated clusters, so describe
some of those results here, as well. More extensive discussions of
most of these issues are contained in our cited works, as well.

\section{The Role of Shocks in Cluster Formation}

Figure 1 illustrates the rich distribution of shock surfaces that are likely
to be associated with large scale structure formation. For
reference a volume rendered image of thermal X-ray emissivity
is included to help locate clusters. These images
are taken from one of our numerical simulations using a grid-based
N-body/hydro scheme (Ryu et al 1993). The hydro part of the code uses a
``TVD'', Riemann-solver scheme that cleanly captures even relatively
weak shocks inside only 2-3 zones, which each span about 100 kpc in the
simulation shown. The volume displayed is a little less than
20 $h^{-1} $Mpc on a side at z = 0, and was extracted from a full simulation box
75 $h^{-1}$ Mpc on a side. For this SCDM model $\Omega_M=1$, $\Omega_B=0.13$,
$h=0.5$ and $\sigma_8 = 0.6$, which produces a cluster population
at the current epoch consistent with observations.
\begin{figure}
\vspace{6.3cm}
\includegraphics{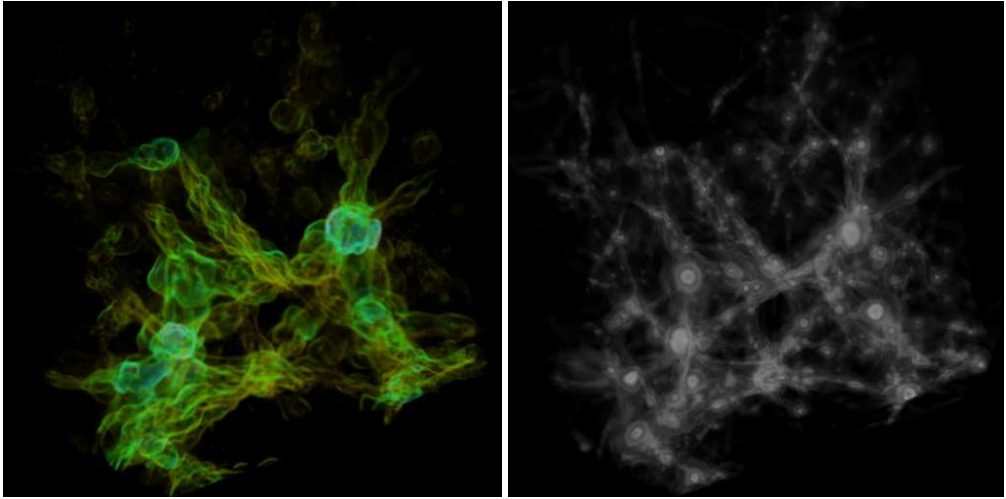}
\caption{Volume renderings of a $(19  h^{-1} Mpc)^3$ portion of 
a $(75 h^{-1} Mpc)^3$ SCDM cosmology simulation showing the locations
of clusters and associated shock structures at $z=0$. Left: Divergence
of the baryonic gas flow, or gas compression rate, filtered to show shocks. 
Right: Bolometric thermal gas emissivity, which peaks in cluster
cores. Note the rich web-like character of the shocks often extending
deeply into the clusters and twisting along the connecting filaments.}
\label{fig1}
\end{figure}

Structure
shocks are most commonly characterized as either ``accretion shocks'', if
they result from infall of diffuse, intergalactic material onto the
perimeter of a cluster, or ``merger shocks'' if they
result from collision between two clusters. A quick glance at
Figure 1 shows that this is an overly simplified picture. For example,
collisions between flows in filaments can lead to shocks, and
accretion shocks are often hard to distinguish from merger shocks, given
the complexity that accompanies the accumulation of mass in regions
where clusters are forming. Dissipation at these shocks provides the basic
heating of the ICM, although other processes, including feedback
from star formation may also be important contributors.

Structure shocks come in a wide range of strengths, most
generally indicated by Mach number. That depends especially on
the temperature of the inflowing gas, since relative flow velocities tend
generally to be of order $10^3$ km $s^{-1}$. To the extent that the interacting
material has been virialized the Mach numbers should be small, of course, 
while cooler material entering a cluster from a filament, for example, may pass through
much higher Mach number shocks. Miniati et al (2000) and Miniati (2002a)
have examined the histories of shock heating as revealed through
cosmological simulations. Integrating over cosmic time they found that 
moderate strength shocks with Mach numbers roughly in
the range $2 \lsim M \lsim 7$ contribute most to heating of the ICM. 
Such shocks are capable of accelerating CRs 
efficiently, so deserve close scrutiny in that context. 

It is important to remember that, since CRs are effectively
tied to the ICM up to pretty high energies,
the integrated shock history of the ICM determines
the character of the CR proton population (as well as their
secondary products). So, one would not expect CRs produced 
in this way to be only
associated with recent merger events, for example. On the other
hand, also keep in mind that electrons lose energy so quickly that we 
can expect to
see them only very close to where they have been accelerated
or, in the case of secondary electrons, produced via decay of
$\pi^{\pm}$. In the context of 
structure shocks, we then should find electronic emissions either in
association with current shocks or as secondaries associated with
the accumulated CR proton population and the thermal baryons.
Synchrotron radiation depends sensitively on magnetic field
strength, so those emissions will also be heavily weighted
towards regions of the strongest fields, wherever they form. 

\section{Diffusive Shock Acceleration}

The DSA  paradigm depends on the ability of
energetic charged particles to pass through the dissipative layer
in a shock mostly unimpeded, but also on the existence of
sufficiently strong scattering to cause the particles to
propagate diffusively upstream and downstream of the shock with
respect to the bulk flow.
That requires a weak magnetic field, which may not exist in 
primordial matter. On the other hand the first stars,
galaxies and perhaps shocks should have seeded the ICM with magnetic
flux. Fields $\gsim 0.1\mu$G should be adequate to accelerate
particles in structure shocks to very high energies 
on reasonable time scales (e.g., Kang et al 1996).
If a shock is planar on the scale of the particle scattering 
lengths and involves a simple, sharp transition, then the
classic ``test particle'' solution for the momentum distribution of
energetic particles is a power law, $f(p) \propto p^{-q}$, independent 
of any other details with a slope given by
\begin{equation}
q = \frac{3r}{r-1}\rightarrow\frac{4M^2}{M^2-1}\approx 4(1~+~\frac{1}{M^2}),
\end{equation}
where $r$ is the compression
ratio of the shock
and the arrow corresponds to a $\gamma = \frac{5}{3}$ gas.
This leads to the well-known behavior, $q \rightarrow 4$, in the strong shock limit. 

For that spectral form the energy and pressure in this population diverge logarithmically
as the momentum, $p \rightarrow \infty$, so it was recognized a long time ago
that diffusive shock acceleration could in principle become
quite efficient (e.g., Axford 1982). Already for $M = 3$, equation 1 gives $q = 4.5$,
while for $M = 5$ the slope is $q = 4.17$.
So, even
moderate shocks can transfer substantial fractions of the kinetic
energy flux into CRs if the seed population is adequate and
the maximum energy is relativistic. The particle acceleration time
scales with the diffusion time across the shock, $t_d = D/u_s^2$.
Again employing a Bohm diffusion model we can estimate that the time to
accelerate CRs to a PeV in a structure formation shock with 
speed $u_s \sim 10^3$km/sec can be as little as $\sim 10^7$yr
when the magnetic field $\sim 1 \mu$G. The seed CR population can come
from previous sources or particles injected
at the shock from the thermal plasma. The latter process is not well understood,
but is certainly nonlinear and
probably depends on the orientation of the magnetic field.
Within our current understanding injection is sometimes modeled as a ``thermal
leakage'' in which a small fraction of the downstream thermalized
population is successful in crossing the postshock
turbulence and the shock itself back into the approaching plasma.
From there it can be returned downstream with an energy boost factor
$\sim u_s/c$ to begin the acceleration process.
Theoretical models (e.g., Malkov 1998) and numerical simulations (Kang et al 2002) suggest
that a simple way to model the injection physics is to match the 
thermal and isotropic nonthermal particle distributions at a particle
speed that is several times the characteristic postshock thermal
particle speed (Kang et al 2002), defining an ``injection momentum''. In the simulations 
described below we will employ that model with injection momentum
set to 2.6 times the postshock thermal peak. Further details
can be found in Miniati 2001. This will generally lead to an injected
fraction of CRs $\sim 10^{-4}$, consistent with more sophisticated
numerical simulations of the diffusive shock acceleration process (Kang et al 2002).

We note as well with this level of CR injection that the
acceleration efficiency of shocks with $M\sim$ 3-5
should be on the order of 10-30\% (e.g., Kang et al 2002).
That estimate is obtained from nonlinear simulations of
CR modified shocks that include feedback between the CR population,
bulk flow upstream and downstream of the shock as well as the
injection process itself. Still, this is sufficiently small
that shock modifications due to CR pressure gradients
are modest and the CR momentum
distribution is not greatly modified from the test particle form.
This is convenient, since it allows us to make reasonable
estimates of CR properties in clusters using test particle models for
acceleration.

\section{Cosmological Simulations Including CR Acceleration at
Shocks}

\subsection{Methods}

A number of authors have explored analytically the acceleration of CRs
by one or more processes (e.g., Kang et al 1996; Eilek \& Wetherall 1999; Blasi 2000;
Brunetti et al 2001)
in clusters. Yet, given the complex
histories of clusters, the absence of stationary structures
and likely wide ranges of properties among individual clusters,
numerical simulations offer a powerful alternative tool to gain
a more complete picture. With this in mind we have
carried out several structure formation simulations
that incorporate an efficient numerical scheme for transport
of CRs, including diffusive acceleration at shocks, as well
as adiabatic losses for protons and radiative and Coulomb
losses for electronic components. The scheme is Eulerian, uses a
finite volume approach to advection in coordinate and momentum space,
and models the momentum distribution as a piecewise power law function
to allow a coarse momentum grid. To
conserve computational effort the simulations described
consider CRs only with energies below 1 PeV, so that
we can reasonably assume acceleration at shocks is 
instantaneous compared to dynamical time steps and that
spatial diffusion of CRs in smooth flows can be ignored. Dynamical
feedback of CRs is neglected; that is, the CR population
is treated in the test particle limit. Thus, CRs
emerging from shocks take a spectral form
governed by equation 1 for injected CRs or whenever that spectrum is flatter
than the incident CR spectrum.

Electrons
are split into a ``primary'' population derived from
thermal leakage at shocks and a ``secondary'' 
population derived from $\pi^{\pm}$ decays following
inelastic p-p collisions. Those same collisions
produce $\gamma$-rays via $\pi^0$ decay, and
that distribution is also computed. Since injection of
primary electrons at shocks is not well understood,
we simply scale it as a fraction, $R_{e/p}$ of the
proton injection described above. Using galactic
CRs for a reference, we may expect $R_{e/p} \sim 0.01$.
The proton, primary and secondary electron populations
are all evolved separately.
Our full numerical method is
an extension of that described in Jones et al (1999)
and has been explained in detail by Miniati (2001). The Eulerian cosmology code
is the same as that employed in the simulation shown in Figure 1,
and outlined in \S 2.

\subsection{Some Results: CR Properties}
\begin{figure}
\vspace{8.5cm}
\includegraphics{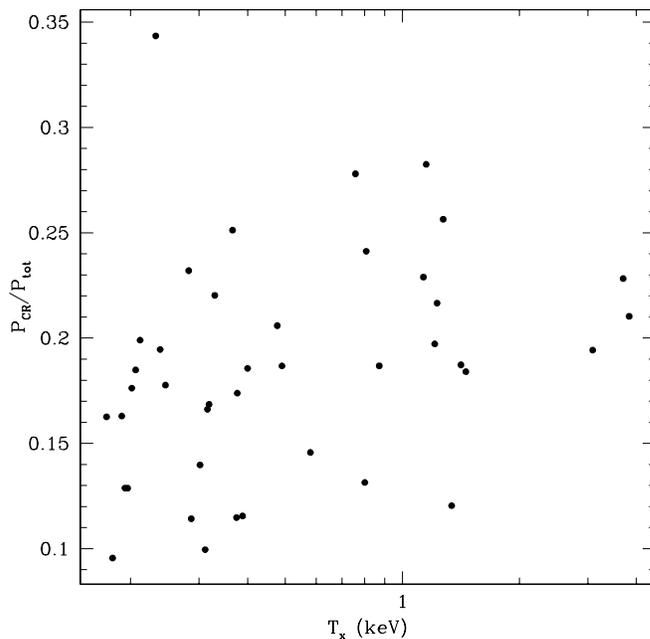}
\caption{ Fractional CR pressure in the central 0.5 $h^{-1}$ Mpc of
individual clusters as a function of cluster core temperature at 
z = 0 in a $\Lambda$CDM simulation as discussed in the text.}
\label{fig2}
\end{figure}

Detailed discussions of results from these
simulations can be found in Miniati et al (2001a,b) and
Miniati (2002a,b). Here we mention briefly a few of
the salient results, especially as they relate to
themes of this meeting.

A single structure formation simulation leads to many clusters
that have formed naturally from random initial fluctuations. While
we pay a price in 
reduced resolution when we simulate a large volume, we gain an unbiased
sample of objects whose properties can be explored from multiple
perspectives.
The simplest question we can ask from these simulations is an
estimate of the total CR population we can expect.
Figure 2 shows estimates of the
fraction of individual cluster core pressures contributed
by CR protons resulting from diffusive acceleration
at structure shocks. These data were obtained from a 
``concordance model'' $\Lambda$CDM simulation with $\Omega_M = 0.3$,
$\Omega_{\Lambda} = 0.7$, $\Omega_B = 0.04$, $h = 0.67$
and $\sigma_8 = 0.9$ (Miniati 2002a,b). (Similar results
were reported for an analogous SCDM simulation, Miniati et al 2001a.) The
simulation shown used a computational box 50 $h^{-1}$ Mpc on a 
side in comoving coordinates with $512^3$ grid zones and
$256^3$ dark matter particles.

The mean CR pressure contribution in these clusters is about 20\% in
the cores, but
there is a very wide scatter. These are first estimates, since
the model for diffusive CR acceleration was applied in the
test particle limit and the injection rates are uncertain. Nonetheless,
the numbers are consistent with expectations from
more sophisticated nonlinear simulations
of individual shocks in the range of Mach numbers mentioned earlier
that are most important for heating of cluster ICMs. 
There is no clear
trend of $P_{CR}/P_{tot}$ with cluster temperature.
The scatter
seen in Figure 2 is real and reflects the influence of varied
shock histories among individual clusters. We emphasize again
that $P_{CR}$
depends on the time integrated shock history of the plasma in
a cluster rather than a single event, such as a recent merger.

Reflecting that fact,
inspection of the spatial distribution of CR protons in the simulations
shows them to be distributed roughly similar to the thermal plasma
(Miniati et al 2001a). However, since the CRs are continuously
injected at accretion
shocks, which lie mostly outside cluster cores, the CR
distributions are broader than the X-ray emitting thermal gas. 
From these behaviors we can immediately anticipate a key property of
secondary electron CRs 
as well as associated $\gamma$-rays from $\pi^0$ decays. Both will be concentrated
in cluster cores, since they will result from a convolution of CR and
thermal proton distributions. That is distinctly different from the distribution of
primary CR electrons and their emissions, which are concentrated
close to contemporary shocks. 
This last point was mentioned earlier
and comes from the very short energy loss time scales of electrons. 
While shocks can and do penetrate into
cluster cores, they are more common, stronger and perhaps more
obvious outside the cores. 

\subsection{Some Results: Nonthermal Emissions}

The spatial and momentum distributions of the CRs can be
used to compute expected emissions from such processes
as inverse-Compton scattering of the CMB, bremsstrahlung
and, since our simulations also include passive magnetic
fields, synchrotron radio emissions. From inelastic p-p scattering
products we include
emissions from secondary $e^{\pm}$ and $\gamma$s from $\pi^0$
decay, as well. 

Since there has been much recent discussion about
nonthermal X-rays and $\gamma$-rays from clusters, we
focus our attentions there.
If the primary electron to proton ratio in cluster CRs
is similar to that in galactic CRs, so that primary
electrons constitute something like a percent of the
total primary CRs, then the average $\gamma$-ray luminosities
in our simulated clusters are produced in comparable
measure by inverse-Compton
emission from primary electrons and $\pi^0$ decay.
The primary electron inverse-Compton contribution is especially
important in extended regions outside cluster cores,
where strong shocks are most likely,
while $\pi^0$ decays can dominate at high energies in cluster cores.

\begin{figure}
\vspace{8.7cm}
\includegraphics{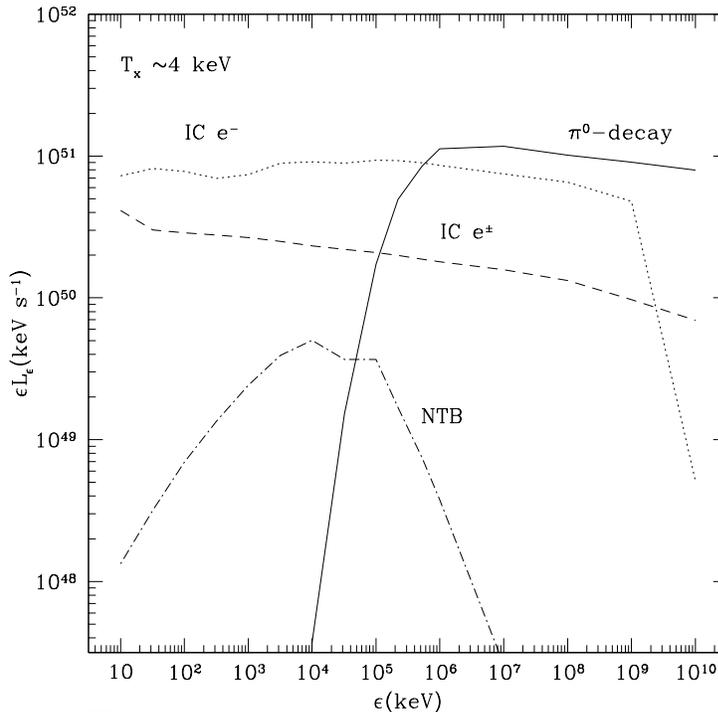}
\caption{X-ray through $\gamma$-ray synthetic spectral energy distribution
for a simulated cluster (adapted from Miniati 2002b).}
\label{fig3}
\end{figure}

Figure 3 illustrates the X-ray to $\gamma$-ray spectral
energy distribution computed from one $T_x \sim 4$ keV cluster 
in the simulation of Miniati (2002a,b). 
IC emission of primary and secondary electrons (actually $e^{\pm}$), 
nonthermal bremsstrahlung
and $\gamma$-rays from $\pi^0$ decay
are included. The emissions were integrated inside a
spherical volume of radius 5 $h^{-1}$ Mpc. 

Primary electron inverse-Compton and $\pi^0$ emissions
dominate as mentioned, while nonthermal bremsstrahlung is largely
unimportant.
Note at high energies that the spectra of inverse-Compton
emissions from both electronic components and $\pi^0$-decay
$\gamma$s are similar over a substantial range. 
All reflect the spectra of primary
protons, with adjustments for radiative losses
in the electrons and for pion generation and decay cross sections in the
other case. Below a GeV, however, the $\pi^0$ contribution
drops due to production threshold and phase space
limits, while above a a few TeV the primary electron
spectrum cuts off since electron energy loss rates
dominate DSA acceleration rates. 

Examination of the spatial distributions of emissions of the
different contributions uncovers some important details
of how structure shocks may reveal themselves, and
suggests some strategies for testing models for
particle acceleration.
Figure 4 separates the emissions from the clusters shown in Figure 3 
into contributions
from the cluster core (2 Mpc diameter) and from the ``outskirts''
defined as an annulus between 1 Mpc and 3 Mpc radius. 
For this plot the cluster was assumed to be at the distance of
the Coma cluster, taken as d~=~105 Mpc.
The figure illustrates
how emissions associated with secondary CRs products concentrate towards
cluster cores, while emissions from from the outer regions
are dominated by primary electrons. Spectral distributions
are distinctly different in the two cases, so it should
be possible in principle to identify emissions from
the hadronic CR component in the cluster cores. That observational
challenge is also currently underway in efforts to
establish the hadronic CR population in galactic supernova
remnants (e.g., Aharonian et al 1994). The right panel in the figure shows the
integrated flux from the different parts of the selected cluster
in comparison to various detection limits.

\begin{figure}
\vspace{7.0cm}
\includegraphics{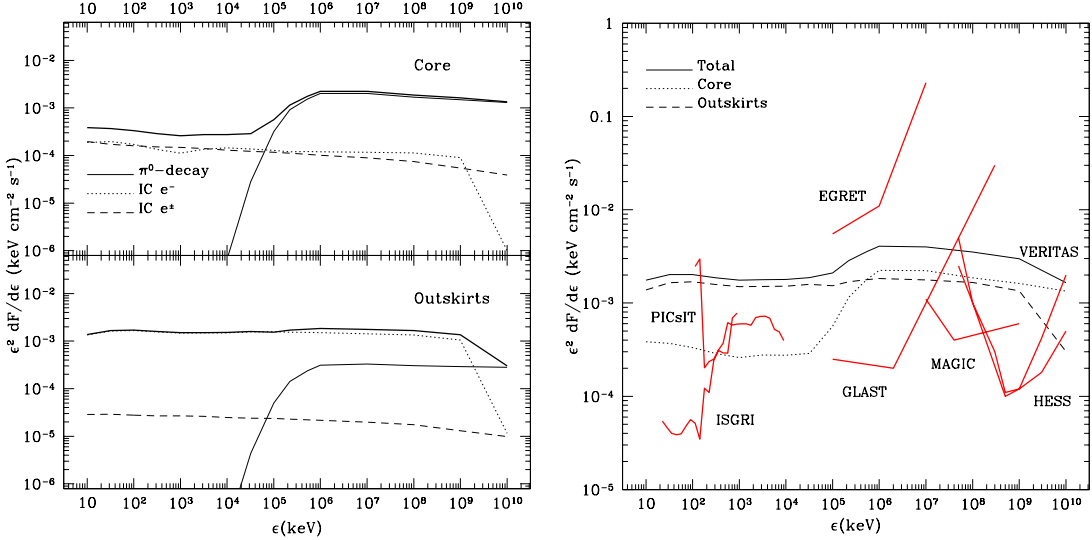}
\caption{Left: Spectral energy distributions from the core (top)
and the outskirts (bottom) of the cluster illustrated in Fig. 3
when the cluster is placed at a distance of 105 Mpc.
The heavy solid lines show the totals in each region. Right:
Total flux compared to various $\gamma$-ray instrument detection limits.
(adapted from Miniati 2002b).}
\label{fig4}
\end{figure}

The above discussion also is relevant to our ability to
understand the origins of diffuse radio halos and
relic sources. Recall that the former are found primarily 
in cluster cores, while the latter tend to lie in
cluster outskirts. In our simulations core regions 
do emit synchrotron emissions that are produced mostly by
secondary electrons. As discussed in Miniati et al 2001b
the radio luminosities and polarization properties
are similar to observed halos, and the
radio luminosity is a very steep function of
cluster temperature, consistent with the
comparative rarity of observed halos. The main handicap in
trying to explain radio halos entirely in terms of 
secondary decay products is that this 
explanation does not appear
to explain naturally observations of radio spectral
steepening (e.g., Deiss et al 1997). However, more sophisticated
models including other contributions to the
electron population or its acceleration 
might solve this problem. The other issue sometimes
raised for secondary electron radio halo models is whether
the required electron population is consistent with
limits to the inverse-Compton X-ray and $\gamma$-ray
fluxes in Coma, for example. For very weak
fields the inverse-Compton flux from the electron
population required to account for radio halo in Coma
would become excessive (Blasi \& Colafrancesco 1999;
Miniati et al 2001a). On the other hand if the
magnetic field is in excess of a few $\mu$G the
required electron population to explain
the synchrotron source is reduced to
a level leading to inverse-Compton luminosities
consistent with current limits
on the $\gamma$-ray flux, and below
the expected $\pi^0$ flux at high energies,
in fact.

\begin{figure}
\vspace{9.0cm}
\includegraphics{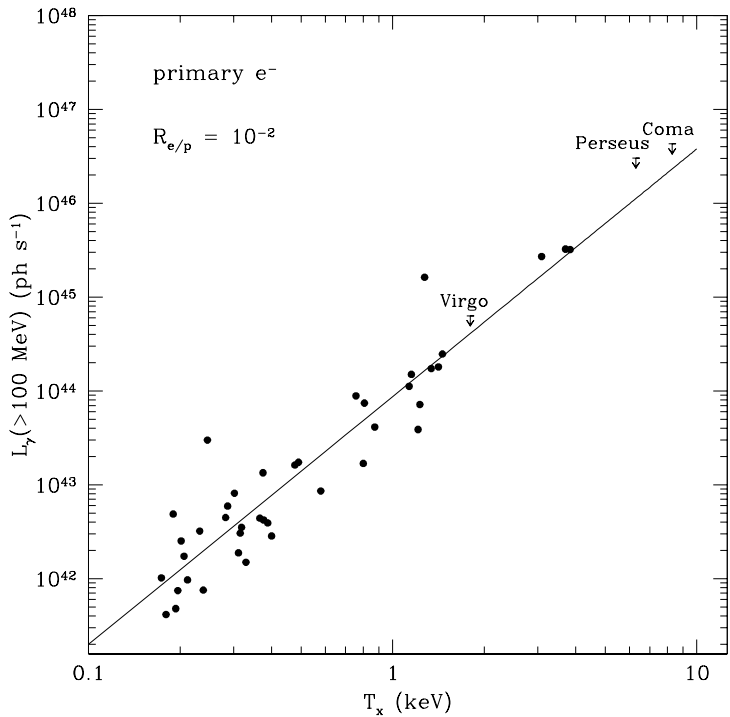}
\caption{ Inverse-Compton $\gamma$-ray luminosity
above 100 MeV from individual clusters as a function of
the cluster core temperature (from Miniati 2002a).}
\label{fig5}
\end{figure}

The simulations also produce diffuse radio sources
resembling radio relics
in some of the same outer regions where the inverse-Compton
emissions from primary electrons are prominent (Miniati et al 2001b).
These are immediate postshock volumes where
primary electrons still remain energetic
and where magnetic fields are relatively strong.
The simulations predict that these regions
should be highly polarized, since the
local magnetic fields tend to become aligned
with the shock faces. These properties
all seem consistent with those seen in
radio relic sources.

Finally we illustrate in
Figure 5 the individual inverse-Compton $\gamma$-ray luminosities
from primary accelerated electrons
computed in a $\Lambda$CDM simulation of
Miniati (2002a,b). 
The solid line is a least
squares power law fit to the simulation data, which has
a slope 2.6, in agreement with what is 
expected from cluster scaling relations (Miniati 2002a). 
The simulated cluster luminosities were integrated
inside a 5 $h^{-1}$ Mpc radius spherical volume. For the
illustration a relative electron shock injection rate of
$R_{e/p} = 0.01$ was used. Computed fluxes scale directly
with that parameter.
Several EGRET upper limits for nearby clusters are shown in
the figure for comparison.
The estimated luminosities from the simulations are consistent with
the current $\gamma$-ray non-detections of clusters, but
they also suggest that $\gamma$-ray fluxes may be large
enough to be seen by the next generation of experiments.

\section{Summary}

Shocks are a ubiquitous consequence of cosmic structure formation,
and they play an essential role in heating of 
cluster media . Virtually all of the gas in
cluster media has been processed by at one or more shocks
of at least moderate strength. Since these shocks involve highly
tenuous ionized media, they are collisionless
in nature, so will not fully thermalize the plasmas
passing through them. One likely consequence is
the acceleration of relativistic particles, or cosmic
rays. We have begun to explore through numerical simulations
the roles that particle acceleration in structure
shocks may play. Our current conclusions are:

$\bullet$The shocks that are primarily responsible
for heating cluster ICMs can be
efficient particle accelerators, possibly
generating nonthermal proton pressures
on the order of 10\% or more of the total 
virial pressure in cluster cores.

$\bullet$Cluster ICMs are very good reservoirs for
energetic protons, so the cosmic ray populations
there reflect the full history of the cluster
medium more than any single event, such as a
recent merger.

$\bullet$Relativistic electrons at most energies
have loss lifetimes so short that they
must either be accelerated locally or
be secondary products from energetic hadronic cosmic
rays in order to have populations
great enough to account for detectable
nonthermal emissions.

$\bullet$There are two main regions for production
of nonthermal radiation in clusters; the X-ray
bright core and the outskirts where strong shocks
are most likely.

$\bullet$Inverse-Compton emission and $\pi^0$
decays dominate the production
of $\gamma$-rays in typical
clusters.  Inverse-Compton from primary
electrons
dominates in the outskirts, provided
electrons are injected in proportion to
ions comparably to the galactic cosmic
rays; that is, so that a
fraction of a percent of the energy flux
through shocks is transferred to 
electrons.

$\bullet$In cluster cores $\gamma$-ray
emission above a few hundred MeV
should be dominated by $\pi^0$ decays,
whereas inverse-Compton emission
from secondary electrons dominates
in these regions at lower $\gamma$-ray 
energies.

$\bullet$Primary and secondary electrons
may also contribute substantially to
nonthermal radio synchrotron emissions
in clusters. Primary electron
emissions are confined to volumes
close to contemporary shocks, so
should be seen mostly in cluster
outskirts, contributing to
radio relic sources. Secondary electronic
emissions should again be 
concentrated in cluster cores, contributing
to radio halos.

\acknowledgments
TWJ and EJH have been supported by 
NASA through grant NAG5-10774, by the
NSF through grant Ast00-71167 and by
the University of Minnesota Supercomputing Institute. 
FM acknowledges support from the Max-Planck-Gesselschaft
Rechenzentrum in Garching.
DR and HK were supported by grant No. R01-1999-00023 from the Korea
Science \& Engineering Foundation.
We thank the organizers of this meeting for their hard work and
for providing a provocative and illuminating program.


\begin{references}

\reference Aharonian, F., A., Drury, L. O'C., \& V\"olk, H. J. 1994, \aap, 285, 645

\reference Axford, W. I., Leer, E., \& McKenzie, J. F. 1982, \aap, 111, 317
\reference Berezinsky, V. S., Blasi, P. \& Ptuskin, V. S. 1997, \apj, 487, 529

\reference Blandford, R. D. \& Eichler, D. 1987, Phys. Rept., 154, 1

\reference Blasi, P. 2000, \apjl, 532, L9

\reference Blasi, P. \& Colafrancesco, S. 1999, Astropart. Phys., 12, 169

\reference Brunetti, G., Feretti, L. \& Giovannini, G. 2001, \mnras, 320, 365

\reference Cen, R. \& Ostriker, J. P. 1999, \apj, 514, 1

\reference Deiss, B. M., Reich, W., Lesch, H., \& Wielebinski 1997, \aap, 321, 55

\reference Dennison, B. 1980, \apjl, 239, L93

\reference Eilek, J. \& Wetherall, J. C. 1999, in
{\it Diffuse Thermal and Relativistic Plasma in Galaxy Clusters}, ed:
H. B\"ohringer, L. Feretti \& P. Schuecker, Garching: MPE, p 249

\reference En$\ss$lin, T. A., Biermann, P. L., Kronberg, P. P. \& Wu, X.-P. 1997, \apj, 477, 560 

\reference Feretti, L. 1999, in 
{\it Diffuse Thermal and Relativistic Plasma in Galaxy Clusters}, ed:
H. B\"ohringer, L. Feretti \& P. Schuecker, Garching: MPE, p 3

\reference Feretti, L. \& Giovannini, G. 1996, in {\it IAU Symposium 175,
Extragalactic Radio Sources}, ed: R. D. Ekers, C. Fanti, \& L. Pardrielli,
Dordrecht: Kluwer, p 333

\reference Fusco-Femiano, R. et al, 2000, \apjl, 534, L7

\reference Jaffe, W. J. 1977, \apj, 212, 1

\reference Jones, T. W., Ryu, D. \& Engel, A. 1999, \apj, 512, 105 

\reference Jones, T. W., Miniati, F., Ryu, D., \& Kang, H. 2001,
in {\it High Energy Gamma-Ray Astronomy}, ed: F. A. Aharonian \& H. J. V\"olk,
New York: AIP, p 436

\reference Kang, H., Ryu, D. \& Jones, T. W. 1996, \apj, 456, 422 

\reference Kang, H., Jones, T. W. \& Gieseler, U. D. J. 2002, \apj (in press)

\reference Malkov, M. A. 1998, Phys. Rev. E, 58, 4911

\reference Miniati, F. 2001, Comp. Phys. Comm, 141, 17

\reference Miniati, F. 2002a, astro-ph/0203014, \mnras (submitted)

\reference Miniati, F. 2002b \mnras (submitted)

\reference Miniati, F., Jones, T. W., Kang, H. \& Ryu, D. 2001b, \apj, 562, 233

\reference Miniati, F., Ryu, Kang, H. \& Jones, T. W. 2001a, \apj, 559, 59

\reference Miniati, F., Ryu, D., Kang, H., Jones, T. W., Cen, R. \&
Ostriker, J. P. 2000, \apj, 542, 608

\reference Quilis, V., Ib\'a\~nez, \& S\'aez, D. 1998, \apj, 502, 518

\reference Rephaeli, Y., Gruber, D. E. \& Blanco, P. 1999, \apjl, 511, L21

\reference Ryu, D., Ostriker, J. P., Kang, H. \& Cen, R. 1993, \apj, 414, 1

\reference Sarazin, C. L. 1999, \apj, 520, 529

\reference V\"olk, H. J.,  Aharonian, F. A., \& Breitschwerdt, D. 1996,  {\it Space Science Reviews}, 75, 279

\end{references}
\end{document}